\documentclass[12pt]{article}
\usepackage{amsmath,amssymb}

\usepackage{graphicx}
\usepackage{wrapfig}

\usepackage{hyperref}
\usepackage{cite}

\def\beq{\begin{eqnarray}}
\def\eeq{\end{eqnarray}}

\def\la{\langle }
\def\ra{\rangle }

\newcommand{\Tr}{\,\mathrm{Tr}\,}            

\newcommand{\eg}{{\it e.g.,}\ }
\newcommand{\ie}{{\it i.e.,}\ }

\newcommand{\be}{\begin{equation}}
\newcommand{\ee}{\end{equation}}
\newcommand{\bea}{\begin{eqnarray}}
\newcommand{\eea}{\end{eqnarray}}
\newcommand{\bg}{\begin{gather}}
\newcommand{\bt}{\beta}
\newcommand{\bseq}{\begin{subequations}}
\newcommand{\eseq}{\end{subequations}}

\renewcommand{\ln}{\mathop{\rm ln}\nolimits}

\def\be{\begin{eqnarray}}
\def\ee{\end{eqnarray}}
\def\lb{\label}

\begin{document}

\title{\textbf{Correlation functions on conical defects}}

\vspace{2cm}
\author{ \textbf{
Michael Smolkin$^\star$ and  Sergey N. Solodukhin$^\sharp$ }} 

\date{}
\maketitle
\begin{center}
\hspace{-0mm}
  \emph{$^{\star}$ Center for Theoretical Physics and Department of Physics}\\
   \emph{  University of California, Berkeley, CA 94720, U.S.A.}
\end{center}
\begin{center}
  \hspace{-0mm}
  \emph{ $^{\sharp}$ Laboratoire de Math\'ematiques et Physique Th\'eorique  CNRS-UMR
7350 }\\
  \emph{F\'ed\'eration Denis Poisson, Universit\'e Fran\c cois-Rabelais Tours,  }\\
  \emph{Parc de Grandmont, 37200 Tours, France} \\
  \emph{ and Theory Group, Physics Department}\\
  \emph{ CERN, CH-1211 Geneva 23,
  Switzerland}
\end{center}

{\vspace{-11cm}
\begin{flushright}
CERN-PH-TH-2014-099
\end{flushright}
\vspace{11cm}
}



\begin{abstract}
\noindent { We explore the new technique developed recently in  \cite{Rosenhaus:2014woa} and suggest a correspondence between the
$N$-point correlation functions on spacetime with conical defects and the $(N+1)$-point correlation functions in regular  Minkowski spacetime.
This correspondence suggests a new systematic way to evaluate the correlation functions on spacetimes with conical defects.
We check the correspondence for the expectation value of a scalar operator and of the energy momentum tensor in a conformal field theory  and obtain the exact agreement with the
earlier  derivations for cosmic string spacetime.  We then use this correspondence and do the computations for a generic scalar operator and a conserved vector current. For generic  unitary field theory we compute the expectation value of the energy momentum tensor using
the known spectral representation of the $2$-point correlators of stress-energy tensor in Minkowski spacetime.
}
\end{abstract}

\vskip 2 cm
\noindent
\rule{7.7 cm}{.5 pt}\\
\noindent 
\noindent
\noindent ~~~ {\footnotesize e-mail: smolkinm@berkeley.edu,  Sergey.Solodukhin@lmpt.univ-tours.fr}

\newpage
    \tableofcontents
\pagebreak

\newpage

\section{ Introduction}
\setcounter{equation}0

In many physics applications we deal with a spacetime which is regular except for a co-dimension two surface 
around which the angle coordinate changes from $0$ to $2\pi\alpha$. Then, for $\alpha$ different from one,
there appears a conical defect. This picture arises in many situations. The most obvious one is the spacetime
created by a cosmic string \cite{Deser:1983tn}. The angle deficit is related to the energy density of the string.
Another similar situation appears in the calculation of the entanglement entropy associated with a
co-dimension two surface $\Sigma$. There, in a replica trick, one allows the angular coordinate 
in the transverse space to $\Sigma$ to have periodicity $2\pi \alpha$. For an integer $\alpha=n$ this corresponds to
gluing together $n$-copies of spacetime. The entropy then is defined by differentiating the partition function
on such $n$-folded spacetime with respect to the angle deficit and by imposing at the end $n=1$. 

Considering various operators and their correlation functions  in all these situations one needs to impose  
the $2\pi \alpha$ periodicity and try not to violate any other nice properties. For free fields Green's functions this is done
by making use of the Sommerfeld formula \cite{Som}.  This allows to achieve the required periodicity of Green's function
and still satisfy the free field equations. For more complicated operators the calculation of correlation functions 
is less simple and often it should be done case by case. 

In this paper we suggest a more regular way of computing the correlation functions on conical defects
by relating them to  the higher order correlation functions defined on a nonsingular Riemannian manifold $\mathcal{M}$. If there is a rotational symmetry around $\Sigma$, then an infinitesimal angular deficit can be treated in a well-defined way following the approach of \cite{Fursaev:1994ea}, where a description of Riemannian geometry in the presence of a conical singularity was studied. For general $\Sigma$ and $\mathcal{M}$  one needs to generalize this method by implementing a squashed cone technique proposed in \cite{Lewkowycz:2013nqa}. As of today the latter technique is not entirely understood, \eg its generalization to the case of gravitational actions that include derivatives of the Riemann tensor is not known. Our general proposal in this paper should hold within the domain of applicability of this generalized method, however we restrict our conclusions to the case of a flat entangling plane embedded in Minkowski space. 

We assume that the theory resides in a vacuum state and define the following operator
\be
{\cal P}=-\lim_{\alpha\rightarrow 1}\, \frac{\partial}{\partial\alpha}
\lb{0}
\ee
which being applied to a function of $\alpha$ extracts a linear term in $(1-\alpha)$. With this definition we suggest the following correspondence 
\be
{\cal P}\la {\cal O}_1(x_1)...{\cal O}_N(x_N)\ra_\alpha=\la {\cal O}_1(x_1)...{\cal O}_N(x_N)K_0\ra_c
\, ,
\lb{2}
\ee
where subscript `c' means connected correlator, ${\cal O}_k$ are arbitrary operators, scalar or tensor, $\la ..\ra_\alpha$ is the correlation function computed in a spacetime with conical defect, $2\pi (1-\alpha)$, localized at $\Sigma$, $\la\cdots\ra$ means vacuum expectation value on $\mathcal{M}$, and $K_0$ is the modular Hamiltonian associated with the vacuum state and a region bounded by $\Sigma$. By definition, the correlator on the left hand side of this correspondence is expanded in powers of $(1-\alpha)$ and only linear term contributes. 

Unfortunately, it is impossible to test our correspondence in full generality since the modular Hamiltonian for generic $\Sigma$ and $\mathcal{M}$ is not known. This operator is not even local in general. However, for certain symmetric geometries, such as spherical and planar regions in Minkowski space, the modular Hamiltonian is expressible in terms of energy-momentum tensor.  As of today, these geometries are probably the only special cases when both sides of the correspondence can be evaluated independently to verify (\ref{2}).  Hence, we first provide a general derivation of the correspondence, and then focus on a very special setup: a plane in a $d$-dimensional Minkowski space. In this setup the entangling surface is symmetric under $O(2)$ rotations around $\Sigma$. We also test (\ref{2}) in the case of a finite interval in a 2D CFT, where $\Sigma$ consists of two disjoint points and therefore lacks rotational symmetry in the transverse space. 

The key ingredient for planar $\Sigma$ is the special role (suggested earlier by many authors \cite{kabastra}) played by operator $K_0$ which generates angular evolution in the  transverse space to $\Sigma$.
This operator is related to the Rindler (or modular) Hamiltonian $H_R$ as $K_0=2\pi H_R$, and it has the following integral representation
\be
K_0=-2\pi\int d^{d-2}y \int_0^\infty dx_1 x_1 T_{22}(x_1,x_2=0,y)\, ,
\lb{1}
\ee
where $(x_1,x_2)$ are Cartesian coordinates in the transverse space, $\Sigma$ is located at the origin
$x_1=x_2=0$ and $y^i$ with $i=1,..,d-2$ are Cartesian coordinates on $\Sigma$. In this notation $x_2$ plays the role of Eulcidean time and $T_{22}$ is the respective component of the energy-momentum tensor.  It is useful to note that the modular Hamiltonian $H_R$ generates angular evolution  in plane $(x_1,x_2)$.

According to our proposal (\ref{2}) the calculation of correlation functions on conical defects now reduces to a calculation in Minkowski
spacetime by inserting a special operator $K_0$. Thus, $N$-point correlation function $\la..\ra_\alpha$ corresponds to an $(N+1)$-point function in Minkowski spacetime. 
This is an improvement over the standard approach since the correlation functions in Minkowski spacetime can be evaluated by using various symmetries  (Poincar\'e and conformal) and  in general are better understood.

That a certain relation between the $N$-point correlation functions on conical defects and higher-point correlation functions in regular flat spacetime
should exist was anticipated in \cite{SS} where it was suggested that the conformal $a$-charge 
normally appearing (in four dimensions) starting with the  $3$-point functions of energy momentum tensor would be visible already 
in $2$-point function considered on spacetime with a defect. Our correspondence (\ref{2}) gives an exact realization of this idea.

It should be noted that after appropriate analytic continuation, the conical spacetime in the case of planar $\Sigma$ in Minkowski space transforms into the Rindler spacetime characterized by some temperature different from the Unruh temperature.
Thus, the correspondence (\ref{2}) should be also valid for the correlation functions in a  thermal field theory  in the Rindler spacetime.

\section{ Derivation of the correspondence}
\setcounter{equation}0

Let us consider a field theory living on a $d$-dimensional Euclidean manifold $\mathcal{M}$ equipped with a Riemanian metric $g_{\mu\nu}$. We assume that the system resides in a slightly excited state given by
\be
 |\Psi\rangle=\exp\big({-g{\cal O}/2}\big)|0\rangle
 \lb{2.1}
\ee
where $|0\rangle$ is the vacuum state of the theory, $g$ is some small dimensionless parameter\footnote{Factor $1/2$ is for later convenience.},
and the scalar operator ${\cal O}$ in general may take the following composite form
\be
{\cal O}=\prod_{k=1}^N \int d^dx_k \, \sqrt{g} \, \sigma_k(x_k)\mathcal{O}_k(x_k) ~ ,
\lb{2.2}
\ee
where $\sigma_k(x_k)$ are arbitrary scalar sources with compact support that couple to operators ${\cal O}_k(x_k)$.

Let us evaluate the entropy in this state for an arbitrary entangling surface $\Sigma$ that divides $\mathcal{M}$ into two subregions $A$ and $B$. The entropy is defined with respect to a reduced density matrix $\rho$ which is obtained from $|\Psi\rangle$ by tracing over the degrees of freedom associated with $A$ 
\be
\rho={\Tr_A |\Psi\rangle\langle\Psi|\over \langle\Psi| \Psi\rangle} ~.
\lb{2.3}
\ee
So that the entropy is
\be
S=-\Tr_B\rho\ln\rho\, .
\lb{2.4}
\ee
The standard way to compute this entropy is to use the replica trick (for a review see \cite{EE})
\be
S=-(\alpha\partial_\alpha-1) \Tr_B\rho^\alpha|_{\alpha=1}\, ,
\lb{2.5}
\ee
where $\Tr_B\rho^\alpha$ is given by a path integral for the theory living on an $\alpha$-folded cover of $\mathcal{M}$ with insertion of $e^{-g{\cal O}}$ on every sheet of the replicated geometry\footnote{In fact, we have insertion of $e^{-g{\cal O}/2}$ on upper ($x_2>0$) and lower ($x_2<0$) parts of each sheet. However, we implicitely assume that $\mathcal{O}_k(x_k)$ are symmetric under parity transformation, and therefore we can combine these insertions into $e^{-g{\cal O}}$.}. In this formulation the entropy to leading order in $g$  is given 
by
\be
S=S_0+g(\alpha\partial_\alpha-1)|_{\alpha=1} \la{\cal O}\ra_\alpha\, ,
\lb{2.6}
\ee
where the expectation value $\la ..\ra_\alpha$ is defined in the `replicated' vacuum state $|0\rangle$ and $S_0$ is the entanglement entropy associated with this state. Note that realization of conifolds in the absence of rotational symmetry around entangling surface require engagement of squashed cone techniques initiated in \cite{Lewkowycz:2013nqa}, and we implicitly assume that this approach is applicable in general. 

Alternatively, the corresponding change in the entanglement entropy can be obtained considering a perturbation in the density matrix, $\rho=\rho_0+\delta\rho$,
where $\rho_0=\Tr_A|0\rangle\langle0|$. To linear order in $\delta\rho$ variation in the entropy is given by the `first law'  of entanglement \cite{Bhattacharya:2012mi}
\be
\delta S=\Tr_B(\delta\rho K_0)\, , \, \, \rho_0=e^{-K_0}\, ,
\lb{2.7}
\ee
here we assumed that the full density matrix is properly normalized, $\text{Tr}_B \rho=\text{Tr}_B \rho_0=1$, so that $\Tr_A\delta\rho=0$. 

If the entangling surface exhibits rotational symmetry in the transverse space, then $K_0$ can be identified with the generator of angular evolution around $\Sigma$, \ie in the case of a plane in flat space it  is given by (\ref{1}). However, there is no closed form expression for the modular Hamiltonian in general. 

In the state (\ref{2.1}) the variation of the density matrix for small $g$ takes the following form
\be
\delta\rho=-g\Tr_A\big( \, |0\ra\la0|{\cal O} \, \big) + g \la 0|\mathcal{O}|0\ra \Tr_A \, |0\ra\la0|~,
\lb{2.8}
\ee
where the last term originates from expansion of $\la\Psi|\Psi\ra$ in (\ref{2.3}) and ensures proper normalization of $\delta\rho$. As a result we get a particular realization of (\ref{2.7}) 
\be
\delta S=-g\Tr_B\Tr_A\big( \, |0\ra\la 0|{\cal O}K_0 \big)+ g \la 0|\mathcal{O}|0\ra \Tr_B \Tr_A\big( \, |0\ra\la 0| K_0 \big)
=-g\la {\cal O}K_0 \ra_c\, ,
\lb{2.9}
\ee
Recall that connected vev on the right hand side is defined on a general Riemanian manifold $\mathcal{M}$. By comparing (\ref{2.9}) with (\ref{2.6}) we arrive at the relation
\be
\lim_{\alpha\rightarrow 1} (\alpha\partial_\alpha-1)\la{\cal O}\ra_\alpha=-\la {\cal O}K_0\ra_c\, .
\lb{2.10}
\ee
In this relation the operator $\cal O$ has the integral form (\ref{2.2}). We note that the integration involves only one power of $\alpha$ despite the
fact that the integral is $N$-multiple. This is due to fact that operator $\cal O$ is inserted only once on each sheet of the replicated geometry.
Taking this remark and  the composite structure of operator (\ref{2.2}) with arbitrary functions\footnote{We assume that the compact supports of $\sigma_k(x_k)$ are disjoint and do not include the singularities of replicated geometry, \ie by assumption non of them overlap with $\Sigma$. To include these cases one needs to account for the standard contact terms as well as possible anomalies that reside on the entangling surface. The latter can be addressed by resorting to Ward identities.} $\sigma_k(x_k)$,   we obtain (\ref{2}) announced in the introduction.

In Minkowski spacetime the expectation value $\la 0| K_0 |0\ra$ vanishes. Therefore we shall  systematically ignore in this paper  the difference between the connected and non-connected correlation functions.

Notice that in the above derivation the scalar operator ${\cal O}_k(x_k)$ is arbitrary, \eg it can be some tensorial operator of the theory contracted with an arbitrary polarization tensor. Furthermore, in the case of flat $\Sigma$ in Minkowski space, operator $K_0$ is defined by (\ref{1}) and therefore the correlation function on the right hand side of (\ref{2}) reduces to an $(N+1)$-correlation function  with insertion of $T_{\mu\nu}$. This type of correlators will be in the focus of explicit computations presented in this paper.

\section{Vacuum expectation  of a scalar operator}
\setcounter{equation}0
\subsection{General consideration}
\lb{sec:scalar}
We start with the  analysis of the simplest case of a 1-point function for a scalar operator $\cal O$ with scaling dimension $\Delta$ in a generic non-conformal field theory. In order to use our 
general prescription (\ref{2}) we need to know a 2-point correlation function $\langle {\cal O}(x) T_{\mu\nu}(x')\rangle$ of this operator  and the energy momentum tensor.
It should be possible to decompose this correlation function in powers of $1/(x-x')^2$ provided this decomposition respects the tensor structure
of this correlation function and the conservation of the energy momentum tensor. These two conditions single out uniquely the following decomposition
\be
&&\langle {\cal O}(x) T_{\mu\nu}(x')\rangle=\sum_{ k} a_k P^{(k)}_{\mu\nu}(x,x')\, ,  \lb{5.1}\\
&&P^{(k)}_{\mu\nu}(x,x')=\frac{1}{(x-x')^{2(d-k)}}\left(\delta_{\mu\nu}-\frac{(d-k)}{(d-2k+1)}\frac{2(x-x')_\mu (x-x')_\nu}{(x-x')^2}\right)\, , \nonumber \\
&&\partial^\mu P^{(k)}_{\mu\nu}(x,x')=0\, ,\nonumber
\ee
where $k\geq(d-\Delta)/2$ is not necessarily an integer. We notice that the conservation law fixes completely the relative coefficient in $P^{(k)}_{\mu\nu}(x,x')$.
Relation (\ref{5.1}) is a small $(x-x')$ expansion which contains both negative and positive powers of $(x-x')$.
The information about concrete scalar operator is now contained in the coefficients $\{a_k\}$.
For $k=\frac{d+1}{2}$ the relevant definition of tensor $P_{\mu\nu}$ is 
\be
P_{\mu\nu}^{(\frac{d+1}{2})}(x,x')=\frac{(x-x')_\mu(x-x')_\nu}{(x-x')^{(d-1)}}\, .
\lb{5.1-1}
\ee

The correlation function  (\ref{5.1}) is $O(d)$-invariant. Now, if our correspondence (\ref{2}) is correct then after two integrations (over $y$ and $x_1$) present in definition (\ref{1}) 
we should end up with a $O(2)$-invariant expression. Let us see how this works for the scalar operator.
We consider the $(22)$ component of (\ref{5.1}), multiply it by $-2\pi$, decompose the coordinates as $x=(y,x_1,x_2)$ and then integrate over $y'$-variables and after that take the
 integration over $x_1'$. The integration over $y'$ is easily performed using the following relation
\be 
&&\int d^{d-2}y'\, P^{(k)}_{22}(y,x_1,x_2;y',x_1',x_2'=0)=\nonumber \\
&&\frac{1}{2}\Omega_{d-3}\frac{\Gamma(\frac{d-2}{2})\Gamma(\frac{d}{2}-k+1)}{\Gamma(d-k)(d-2k)(d-2k+1)}\, \frac{\partial^2}{\partial^2 {x'_1}}\left[\frac{1}{(x_2^2+(x_1'-x_1)^2)^\frac{{d-2k}}{2}}\right]\, ,
 \lb{5.2}
 \ee
where
$$
\Omega_{d-3}=\frac{2\pi^{\frac{d-2}{2}}}{\Gamma(\frac{d-2}{2})}
$$
is the area of unite $(d-3)$-sphere. We stress that namely due to the precise balance between two terms in the expression for $P^{(k)}_{\mu\nu}$
we have this second derivative form for the integral (\ref{5.2}) which otherwise would not have happened. 

There are two special cases when we have to re-evaluate this integral more carefully. The first case is when $k=\frac{d+1}{2}$. Then we have
\be
\int d^{d-2} y'\, P^{(\frac{d+1}{2})}_{22}(y,x_1,x_2;y',x_1',x_2'=0)=
\frac{\sqrt{\pi}}{4}\Omega_{d-3}\frac{\Gamma(\frac{d}{2}-1)}{\Gamma(\frac{d+1}{2})}\frac{\partial^2}{\partial^2 {x'_1}}\sqrt{x_2^2+(x_1'-x_1)^2}\, .
\lb{5.2-1}
\ee
The second special case is when $k=\frac{d}{2}$.  In this case the integration in (\ref{5.2}) gives
\be
\int d^{d-2}y'P^{(\frac{d}{2})}_{22}(y,x_1,x_2;y',x_1',x_2'=0)=-\frac{\Omega_{d-3}}{2(d-2)}\frac{\partial^2}{\partial^2 {x'_1}} \ln(x_2^2+(x'_1-x_1)^2)\, .
\lb{5.2-3}
\ee

Now the integration  remained  over $x_1'$ is performed as follows
\be
\int_0^\infty dx'_1x_1'\frac{\partial^2}{\partial^2 {x'_1}}\left[\frac{1}{(x_2^2+(x_1'-x_1)^2)^\frac{{d-2k}}{2}}\right]=\frac{1}{(x_1^2+x_2^2)^{d-2k\over 2}}\, .
\lb{5.3}
\ee
For values of $n$ such that $(2k-d)>0$ we use a regularization which consists in evaluating the integral only at lower limit of integration $x_1'=0$ and ignoring the terms coming from the infinity. Technically this is done by evaluating the integral in (\ref{5.3}) first for $d-2n>0$ and then taking the continuous limit to values of $n$ such that $d-2n<0$.

Collecting everything together and using our correspondence  we   arrive at our result
\begin{eqnarray}
&&{\cal P}\langle {\cal O}(y,x_1,x_2)\rangle_\alpha=-\pi\Omega_{d-3}\sum_{k} \frac{\Gamma(\frac{d-2}{2})\Gamma(\frac{d}{2}-k+1)}{\Gamma(d-k)(d-2k)(d-2k+1)}\, \frac{a_k}{r^{d-2k}}
\nonumber \\
 &&-  \frac{\pi^{3/2}}{2} \frac{\Gamma({d-2\over 2}) }{\Gamma(\frac{d+1}{2})} \, \Omega_{d-3}\, a_{{d+1\over 2}} \, r + {2\pi\over (d-2)}\,\Omega_{d-3}\, a_{\frac{d}{2}}\, \ln r  ~,
\lb{5.4}
\end{eqnarray}
where $r=\sqrt{x_1^2+x_2^2}$ and in the second line we explicitly included two special terms with $k=d/2$ and $k=(d+1)/2$ respectively. We see that the resulting correlation function does not depend on coordinate $y$ on the singular surface $\Sigma$. This is as expected since
we have a translational symmetry along the surface. On the other hand, the correlation function is $O(2)$ invariant in the direction orthogonal to the surface and thus it depends only on the distance $r$ to the singularity. Formally, this can be seen as follows. The modular Hamiltonian is a generator of  rotations in plane $(x_1,x_2)$ and hence one has that
$2\pi\partial_\varphi {\cal O}=[{\cal O},\, K_0]$. Therefore, the expectation value $\partial_\varphi \la 0|{\cal O}K_0|0\ra=0$ as we observe in a direct calculation in (\ref{5.4}).

By our proposal (\ref{2}), equation (\ref{5.4}) represents  a linear in $(1-\alpha)$ term in the expansion of expectation value of $\cal O$ evaluated on a conifold.
In order to test our proposal we now consider a particular example of a minimally coupled scalar field.

\subsection{Free massive scalar field}
In this subsection we shall evaluate the expectation value of $\phi^2$ on a conical defect using our correspondence and then compare this with the known results.
Let us first consider a minimally coupled scalar field of mass $m$. The corresponding energy momentum tensor is the canonical one
\be
 T^c_{\mu\nu}=\partial_\mu\phi\partial_\nu\phi-\delta_{\mu\nu}\left( {1\over 2} (\partial\phi)^2+{m^2\over 2}\phi^2 \right)\, .
 \lb{5.2.1}
\ee
We are interested to evaluate the following connected correlator
\be
&&\langle \phi^2(x) T^c_{\mu\nu}(x')\rangle=2\langle \phi(x)\partial_\mu\phi(x')\rangle\langle \phi(x)\partial_\nu\phi(x')\rangle\nonumber \\
&&-\delta_{\mu\nu}\bigg( \langle \phi(x)\partial_\mu\phi(x')\rangle\langle \phi(x)\partial^\mu\phi(x')\rangle + m^2 \langle\phi(x)\phi(x')\rangle^2 \bigg)~.
 \lb{phi2T}
\ee
The Euclidean two-point function in free field has the standard form 
\be
 \langle\phi(x)\phi(x')\rangle=\int {d^dp\over (2\pi)^d} {e^{ip\cdot (x-x')} \over p^2+m^2}={1\over 2\pi} \left( {m\over 2\pi \sigma} \right)^{d-2\over 2} K_{d-2\over 2}(m \sigma)\, ,
 \lb{prop}
\ee
where $\sigma=\sqrt{(x-x')^2}$. Substituting this   into (\ref{phi2T}) yields
\be
&& \langle \phi^2(x) T^c_{\mu\nu}(x')\rangle\nonumber \\ 
&&=- {1\over (2\pi)^d} {m^d\over  \sigma^{d-2}} \left( \delta_{\mu\nu}\Big(K_{d\over 2}^2(m \sigma)+K_{d-2\over 2}^2(m \sigma) \Big)
 -2{(x-x')_\mu (x-x')_\nu\over \sigma^2}K_{d\over 2}^2(m \sigma) \right)~.
\ee
In order to make contact with our general representation  (\ref{5.1}) we  expand this expression in  $m$ or, equivalently, in powers of $\sigma$ using
the expansion formula for the Bessel function
\be
 K_{d\over 2}^2(x)=x^{-d} \, \Gamma(d/2) \, \left( 2^{d-2}\Gamma(d/2) - 2^{d-3}\Gamma\left(d/2-1\right) x^2 + \ldots  \right)\, .
 \lb{expK}
\ee
We then find
\be
\langle \phi^2(x) T^c_{\mu\nu}(x')\rangle= - {\Gamma(d/2)^2\over 4\pi^d}\left(P^{(1)}_{\mu\nu}(x,x')-m^2\frac{(d-3)}{(d-2)^2}P^{(2)}_{\mu\nu}(x,x')+..\right)\, 
\lb{OK}
\ee
in terms of $P^{(n)}_{\mu\nu}$ introduced in (\ref{5.1}).
Now we can use our general result (\ref{5.4}) and, after some simplifications, arrive at the following expression
\be
{\cal P}\langle\phi^2(x)\rangle_\alpha={\Gamma(d/2)^2\over2^d\pi^{d-1\over 2}(d-2)\Gamma\left({d+1\over2}\right)} {1\over r^{d-2}}\left(1- {2(d-1)\over (d-2)(d-4)}(mr)^2+\ldots\right)\, ,
\lb{phi2K}
\ee
where $r=\sqrt{x_1^2+x_2^2}$ is the distance to the conical singularity.

In $d=4$ the second term in (\ref{OK}) is the one which corresponds to $n=d/2$. It should be treated separately using (\ref{5.2-1}) and it produces the logarithmic term outlined in (\ref{5.4}). 
With this term taken into account the result in four dimensions is as follows        
\be
{\cal P}\langle\phi^2(x)\rangle_\alpha={1\over 24\pi^2 r^2}\big(1 +  3 (mr)^2 \log(mr)+\ldots\big)\, .
\ee
This result can now be compared with those available in the literature, see for instance \cite{Iellici:1997ud}, for the direct calculation in conical spacetime and we find exact agreement. This is an important  check for our general proposal. We however stress that our results in this section are much more general. They are valid for any
scalar operator in arbitrary dimensions.

It is straightforward to generalize our discussion to include a non minimally coupled scalar field. In this case (\ref{5.2.1}) undergoes the following improvement
\be
 \tilde T_{\mu\nu}=T^c_{\mu\nu}+\xi(\delta_{\mu\nu}\partial^2-\partial_\mu\partial_\nu)\phi^2\, .
\lb{5.3.1}
\ee
For conformally coupled scalar 
\be
\xi_{c}={d-2\over 4(d-1)}~.
\ee
New term in the energy-momentum tensor induces the following correction to (\ref{phi2T})
\be
&&\langle \phi^2(x)  \tilde T_{\mu\nu}(x')\rangle=\langle \phi^2(x) T^c_{\mu\nu}(x')\rangle
+\xi(\delta_{\mu\nu}\partial^2-\partial_\mu\partial_\nu) \langle\phi(x)\phi(x')\rangle^2 ~,
\lb{5.3.2}
\ee
where derivatives act on $x'$.

Now using (\ref{prop}) and expansion (\ref{expK}) one can carry out all necessary integrals as we did it in section \ref{sec:scalar}. The final result reads
\begin{multline}
{\cal P}\langle\phi^2(x)\rangle_\alpha={\Gamma\big({d-2\over 2}\big)^2\over (4\pi)^{d-1\over 2}\Gamma\left({d-1\over2}\right)} \,{1\over r^{d-2}}
\left(\xi_c-\xi 
+ {2(d-3)\over (d-4)^2}\bigg(\xi - {d-4\over 4(d-3)} \bigg)(mr)^2+\ldots\right)\, .
\lb{5.3.3}
\end{multline}
In the conformal case $\xi=\xi_c$, $m=0$ this expression vanishes.
It should be mentioned that this result is expected. Indeed, by conformal invariance the correlation function $\la {\cal O}(x)T^{CFT}_{\mu\nu}(x')\ra$
vanishes identically and so does the correlation function of ${\cal O}$ with operator $K_0$. This happens because the correlation function of ${\cal O}$ with the improvement
in  (\ref{5.3.1}) cancels the correlation function with the canonical energy momentum tensor.

\section{ Vacuum expectation of a conserved vector current}
\setcounter{equation}0
We can generalize our consideration of the previous section for the case when the operator which appears in the perturbation (\ref{2.1})
is a conserved vector current $J_\mu(x)$, $\partial^\mu J_\mu(x)=0$. The correlation function of this vector current and the energy momentum tensor 
in Minkowski spacetime 
\be
\la J_\mu(x)T_{\alpha\beta}(x')\ra =P_{\mu,\alpha\beta}(x,x')\, 
\lb{c1}
\ee
in general is a combination of terms 
\be
&&P_{\mu,\alpha\beta}(x,x')=\frac{1}{{(x-x')}^{2k}}(A\delta_{\alpha\beta}(x-x')_\mu+B(\delta_{\alpha\mu}(x-x')_\beta+\delta_{\beta\mu}(x-x')_\alpha) 
 \nonumber \\
&& +C\frac{(x-x')_\mu(x-x')_\alpha (x-x')_\beta}{(x-x')^2})
\lb{c2}
\ee
provided the conservation laws 
\be
\partial^\mu P_{\mu,\alpha\beta}(x,x')=\partial^{\alpha}P_{\mu,\alpha\beta}(x,x')=0\, 
\lb{c3}
\ee
are satisfied.
These conditions happen to be very restrictive. They fix not only the possible values of constants $A,\, B,\, C$ but also the power $k$.
In fact, we have found only two possible solutions to these conditions
\be
&&P^{(I)}_{\mu,\alpha\beta}(x,x')=\frac{(x-x')_\mu (x-x')_\alpha (x-x')_\beta}{(x'-x)^{d+2}}\, \ , \nonumber \\
&&P^{(II)}_{\mu,\alpha\beta}(x,x')=\frac{1}{(x'-x)^{d+2}}\bigg(\frac{(x'-x)_\mu (x'-x)_\alpha (x'-x)_\beta}{(x'-x)^2}\nonumber \\
&&-\frac{1}{(d+2)}\big(\delta_{\alpha\beta}(x'-x)_\mu+
\delta_{\alpha\mu}(x'-x)_\beta+\delta_{\beta\mu}(x'-x)_\alpha\big)\bigg)\, .
\lb{c4}
\ee
An interesting property of the solution of type $II$ is that it is traceless with respect to any pair of indexes, $\delta^{\alpha\beta}P^{(II)}_{\mu,\alpha\beta}=\delta^{\alpha\mu}P^{(II)}_{\mu,\alpha\beta}=0$.

Respectively, we have two types of conserved currents which we shall call $J^{(I)}_\mu(x)$ and $J^{(II)}_\mu(x)$.
In order to compute the expectation value of these currents on a conical defect we shall use  our correspondence 
along the same lines as we have done this for a scalar operator in the previous section. 
The results of the calculation can be summarized as follows.

For the vector current of type I we find the expectation values
\be
&&{\cal P}\la J^{(I)}_i(y,x_1,x_2)\ra_\alpha=0\, , \, \,  i\in\Sigma ~, \nonumber \\
&&{\cal P}\la J^{(I)}_1(y,x_1,x_2)\ra_\alpha=-\frac{2\pi \, \Omega_{d-3}}{d(d-2)} \, x_2\Big(\frac{\pi}{2}+\arctan\frac{x_1}{x_2}\Big)\, , \nonumber \\
&&{\cal P}\la J^{(I)}_2(y,x_1,x_2)\ra_\alpha=-\frac{2\pi \, \Omega_{d-3}}{d(d-2)}\, x_2\Big(1+\frac{x_1}{x_2}\Big(\frac{\pi}{2}+\arctan\frac{x_1}{x_2}\Big)\Big)~.
\lb{c5}
\ee
For the vector current of the type II we find the expectation values 
\be
&&{\cal P}\la J^{(II)}_i(y,x_1,x_2)\ra_\alpha=0\, , \, \,  i\in\Sigma ~,\nonumber \\
&&{\cal P}\la J^{(II)}_a(y,x_1,x_2)\ra_\alpha=-\frac{2\pi \, \Omega_{d-3}}{d(d-2)(d+2)}\frac{x_a}{r^2}\, , \, \, a=1,2\, ,
\lb{c6}
\ee
where $r=\sqrt{x_1^2+x_2^2}$. We see that in both cases the only non-vanishing components of $\la J_\mu\ra$ are those lying in the plane perpendicular to the surface $\Sigma$
so that the vector current in this plane  are divergence free,
\be
\partial^a \la J^{(I,II)}_a\ra_\alpha =0\, .
\lb{c7}
\ee
The result (\ref{c5}) for the current $I$ is somewhat puzzling. It is not obviously rotational invariant and, moreover, it is multiple valued function of angular coordinate
in the plane $(x_1,x_2)$. 
It would be nice to have a better understanding of this behavior and compare these results with the direct calculations for vector currents in spacetime of cosmic string.
We, however, at the moment are not aware of any such calculations.

\section{ Vacuum expectation of $ T_{\mu\nu}$ in $CFT_4$}
\setcounter{equation}0
In order to further  check our proposed correspondence (\ref{2}) we shall consider a conformal field theory in four dimensions and 
compute the expectation value of the stress energy tensor in this theory and compare it with the known results in the literature
obtained directly on a conical spacetime.

Conformal symmetry as is well known \cite{Osborn:1993cr} fixes completely the structure of 2-point function for the energy momentum tensor,
\begin{equation}
\langle T_{\mu \nu}(x) T_{\alpha\bt}(x')\rangle = \frac{C_T \, \mathcal{I}_{\,\mu \nu, \alpha\bt}} {\left((x-x')^2 \right)^4}
\lb{3.1}
\end{equation}
with 
\be
&&\mathcal{I}_{\mu \nu, \alpha\bt} = {1\over 2} \left( I_{\mu \alpha}I_{\nu \bt} + I_{\mu \bt}I_{\nu \alpha} \right)- \frac{1}{4} \delta_{\mu \nu}\delta_{\alpha\beta} ~,\nonumber \\
&&I_{\mu \nu} = \delta_{\mu \nu} - 2\frac{ (x-x')_{\mu} (x-x')_\nu}{(x-x')^2}~.
\lb{3.2}
\ee
$C_T$ is a charge related to the $B$-type conformal anomaly.
As is seen from the structure of the operator $K_0$ (\ref{1}) we shall need only component $\alpha=\beta=2$ of (\ref{3.1})
in order to compute correlator $\la T_{\mu\nu}(x)K_0\ra$. The integration over $y$-variables and over $x_1$ is rather straightforward and we find that
\be
&&\la T_{ij}(x_1,x_2,y)K_0\ra=\frac{\pi^2}{120}C_T \frac{\delta_{ij}}{r^4}\, , \, \, i,j=3,4 \nonumber \\
&&\la T_{ab}(x_1,x_2,y)K_0\ra=\frac{\pi^2}{120}C_T \frac{4x_ax_b-3\delta_{ab}r^2}{r^6}\, ,\, \, a,b=1,2
\lb{3.3}
\ee
where $r^2=x_1^2+x_2^2$.
These correlation functions do not depend on the coordinate $y$ as a consequence of the translation invariance along the planar surface $\Sigma$.
For a free field multiplet containing  $n_s$ conformal scalars, $n_f$ Dirac fermions and $n_v$ gauge vector fields we have
\be
C_T=\frac{1}{3\pi^4}(n_s+6n_f+12n_v)\, .
\lb{3.4}
\ee
The result (\ref{3.3}) should be compared to the well known result \cite{cosmic} for the expectation value of energy momentum tensor computed directly on a conical spacetime.
Written in  polar coordinates it takes the form 
\be
 \langle T^{\mu}_{\nu} \rangle_\alpha={f(\alpha)\over 1440 \pi^2 r^4} \, \text{diag}(1,-3,1,1)~,
 \lb{3.5}
\ee
where 
\be
 f(\alpha)={1\over 16 \alpha^4}(1-\alpha^2)\big[2\big(1+{\alpha^2\over 3}\big)a+\big(1-\alpha^2\big)\big(2b-c\big)\big]~,
\ee
with
\be
&&a=12(n_s+6n_f+12n_v)\, , \nonumber \\
&&b=-4(n_s+\frac{11}{2}n_f+62n_v)\, , 
\nonumber \\
&&c=-240n_v\, .
\ee
Applying now operator ${\cal P}$ (\ref{0}) to correlation function
(\ref{3.5}) we find a complete agreement with our result (\ref{3.3}). This is a rather non-trivial check on our proposed
correspondence.

\section{ Vacuum expectation of $ T_{\mu\nu}$  in a generic $d$-dimensional field theory}
\setcounter{equation}0

The result of the previous section can be generalized to any unitary (non necessarily free) field theory by making use of the spectral representation of the
two-point correlation function of energy momentum tensor. This representation was suggested in \cite{Cappelli:1990yc} and it takes the form
\be
 \langle T_{\alpha\beta}(x) T_{\rho\sigma}(x') \rangle &=& {A_d \over (d-1)^2} \int_0^\infty d\mu ~ c^{(0)}(\mu) ~ \Pi^{(0)}_{\alpha\beta,\rho\sigma}(\partial)~G_d(x-x',\mu) 
 \nonumber \\
 &+&{A_d \over (d-1)^2} \int_0^\infty d\mu ~ c^{(2)}(\mu) ~ \Pi^{(2)}_{\alpha\beta,\rho\sigma}(\partial)~G_d(x-x',\mu) 
  ~,
\lb{4.1}
\ee
where $A_d={\Omega_{d-1}\over (d+1) 2^{d-1}}~, \ \Omega_{d-1}={2\pi^{d/2}\over \Gamma (d/2)}$ and we defined operators
\be
 &&\Pi^{(0)}_{\alpha\beta,\rho\sigma}(\partial)= {1\over \Gamma(d)} S_{\alpha\beta} S_{\rho\sigma}\, ,
 \nonumber \\
 &&\Pi^{(2)}_{\alpha\beta,\rho\sigma}(\partial)= {d-1\over 2\,\Gamma(d-1)}\left( S_{\alpha\rho}S_{\beta\sigma}+S_{\alpha\sigma}S_{\beta\rho} - {2\over d-1}S_{\alpha\beta} S_{\rho\sigma} \right)~,
 \lb{4.2}
\ee
whereas $S_{\alpha\beta}=\partial_\alpha\partial_\beta- \delta_{\alpha\beta}\partial^2$ and
\be
 G_d(x-x', \mu)=\int {d^dp \over (2\pi)^d} {e^{ip\cdot(x-x')}\over p^2+\mu^2}~
 \lb{4.3}
\ee
is Green's function of a massive scalar field in $d$-dimensions. The representation (\ref{4.1}) is general and it is valid for any unitary theory.
The information about the concrete quantum field theory is encoded in the spectral function $c^{(0)}(\mu)$ and $c^{(2)}(\mu)$.

This spectral representation of the correlation function is helpful  and we can now compute the correlation function of the product of the energy momentum tensor
and operator $K_0$,
\be
&&\la T_{\alpha\beta}(x_1,x_2,y)K_0\ra=
-{2\pi A_d \over (d-1)^2}\int_0^\infty d\mu(c_{0}(\mu)\Pi^{(0)}_{\alpha\beta,22}(\partial)+c_{(2)}(\mu)\Pi^{(2)}_{\alpha\beta,22}(\partial))\nonumber \\
&&\int d^{d-2}y'\int_0^\infty dx'_1 x_1'G_d(x_1-x_1',x_2,y-y',\mu)\, .
\lb{4.4}
\ee
First we notice that the integration over $y'$ in (\ref{4.4}) produces a two-dimensional Green's function,
\be
\int d^{d-2}y'G_d(x_1-x_1', x_2,y-y',\mu)=G_{2}(x_1-x_1',x_2,\mu)=\frac{1}{2\pi}K_0(\mu\sqrt{(x_1-x_1')^2+x_2^2}) \, .
\lb{4.5}
\ee
In fact, it is not surprising that we get a two-dimensional Green's function. This is a well known method, called the {\it descent method}, to get a lower dimensional Green's function from a higher-dimensional one
by integrating over a subset of variables.

The function (\ref{4.5}) does not depend on coordinates $y$. Therefore the derivatives with respect to $y$ in (\ref{4.4}) give the vanishing results when act on (\ref{4.5}).
This indicates that the operators $S_{\alpha\beta}$ reduce to purely two-dimensional operators acting in the transverse sub-space,
\be
 S_{\alpha\beta} \rightarrow 
 \left\{\begin{array}{c} \partial_a\partial_b-\delta_{ab}\,\Delta^{(2)} \quad \text{for}~ \alpha,\beta=a,b=1,2~,\\ -\delta_{ij}\,\Delta^{(2)}  ~\quad\quad\quad \text{for}~ \alpha,\beta=i,j ~, \end{array}\right.
\ee
where $\Delta^{(2)}=\partial^2_1+\partial^2_2$ is the two-dimensional Laplacian in the transverse space to surface $\Sigma$.
Due to this the operators ({\ref{4.2}) are simplified
\be
&&\Pi^{(0)}_{ij,22}=\frac{\delta_{ij}}{\Gamma(d)}\Delta^{(2)}\partial_1^2\, ,\, \, \Pi^{(2)}_{ij,22}=-\frac{\delta_{ij}}{\Gamma(d-1)}\Delta^{(2)}\partial_1^2\, , \nonumber \\
&&\Pi^{(0)}_{ab,22}=-\frac{1}{\Gamma(d)}S_{ab}\partial^2_1\, , \, \, \Pi^{(2)}_{ab,22}=-\frac{d-2}{\Gamma(d-1)}S_{ab}\partial^2_1\, .
\lb{4.6}
\ee
On the other hand, since $G_2$ is a solution to equation $\Delta^{(2)}G_2=\mu^2G_2$  one can replace $\Delta^{(2)}$ by $\mu^2$.
What is left is the integration over $x'_1$ in (\ref{4.4}). This can be easily done by integrating twice by parts and we arrive at a simple expression
\be
\int_0^\infty dx'_1x'_1\frac{\partial^2}{\partial x_1^2}G_2(x_1-x'_1,x_2,\mu)=G_2(x_1,x_2,\mu)=\frac{1}{2\pi}K_0(\mu \sqrt{x_1^2+x_2^2})\, .
\lb{4.7}
\ee
We pause here to appreciate this little magic. Indeed, we have started with expression (\ref{4.4}) in which $O(2)$ symmetry in $(x_1,x_2)$ subspace was not evident at all
but then after performing all integrations we arrive at an expression which is clearly $O(2)$ invariant. From the conical space point of view this is of course expected due to the rotational symmetry  around the tip of the cone.

The final result for the correlation function (\ref{4.4}) can be now represented in a form of the integral over  the spectral parameter $\mu$,
\be
&& \langle T_{ij}(x_1,x_2,y)  K_0 \rangle = -{A_d \, \delta_{ij}\over (d-1)^2 \Gamma(d)} \int_0^\infty d\mu ~ \big(c^{(0)}(\mu) -(d-1)c^{(2)}(\mu) \big)
 \, \mu^2 K_0(\mu\, r)\, ,\nonumber \\ 
&& \langle T_{ab}(x_1,x_2,y)  K_0 \rangle=-{A_d \over (d-1)^2 \Gamma(d)} \int_0^\infty d\mu ~ \big(c^{(0)}(\mu) +(d-1)(d-2)c^{(2)}(\mu) \big)\nonumber \\
&&\left(\mu^2K_0(\mu r)(\delta_{ab}-\frac{x_ax_b}{r^2})+\frac{\mu K_1(\mu r)}{r}(\delta_{ab}-\frac{2x_ax_b}{r^2})\right)\, .
\lb{4.9}
\ee
This represents the general result valid for arbitrary (free or interacting) field theory. By our proposed correspondence (\ref{2}) this expression is our prediction for the   leading order contribution to the expectation value of energy momentum tensor on a planar conical defect in such a generic  $d$-dimensional theory.

\medskip

We now consider some particular case.

\bigskip

\noindent{\it Conformal field  theory limit.}
\medskip
Conformal field theory is a particular example for the theory considered above. 
As argued in \cite{Cappelli:1990yc}, in this case  we have 
\be
 c^{(0)}(\mu)\overset{CFT}{\propto} \mu^{d-2}\delta(\mu)~, \quad c^{(2)}(\mu)\overset{CFT}{=}{d-1\over d} C_T \, \mu^{d-3}~,
 \lb{4.10}
\ee
where $C_T$ is the charge which appears in the 2-point correlation function of energy momentum tensor in $d$-dimensional CFT,  similarly to 
the four-dimensional case (\ref{3.1}). Performing the integration over $\mu$ we obtain
\be
&& {\cal P}\la T_{ij}(x_1,x_2,y)\ra_\alpha=\langle T_{ij}(x_1,x_2,y)  K_0 \rangle\overset{CFT}{=}\, \frac{C_T\pi^{d/2}\Gamma(\frac{d}{2})}{\Gamma(d+2)}\frac{\delta_{ij}}{r^d}\, ,\, \, i,j=3,..,d \lb{4.11}\\
&& {\cal P}\la T_{ab}(x_1,x_2,y)\ra_\alpha= \langle T_{ab}(x_1,x_2,y)  K_0 \rangle \overset{CFT}{=}\, \frac{C_T\pi^{d/2}\Gamma(\frac{d}{2})}{\Gamma(d+2)}\left(d\,\frac{x_ax_b}{r^2}-(d-1)\delta_{ab}\right)\, , \, \, a,b=1,2\, \nonumber
\ee
In four dimensions we again reproduce (\ref{3.3}). We are not aware of any previous results in higher dimensions. Therefore, (\ref{4.11})  is our prediction for the expectation value of a CFT energy momentum tensor
on a conical defect.

\section{Finite interval in a 2D CFT}
\setcounter{equation}0

In this section we want to test our correspondence for more general geometries. Let us consider a two dimensional conformal field theory living in $\mathbb{R}^2$. By assumption, the theory resides in a vacuum state and we choose some interval of finite length $\ell$ to represent sub-system $B$. This setup is probably a simplest example where $\Sigma$ is not rotationally symmetric in the transverse space and still both sides of the correspondence can be explicitly evaluated. 

We parametrize $\alpha$-folded cover of $\mathbb{R}^2$ by $w$ and denote the end points of the interval by $u$ and $v$. Now the conformal transformation $z=((w-u)/(w-v))^{1/\alpha}$ maps the $\alpha$-sheeeted Riemann surface to the ordinary complex plane $\mathbb{C}$  \cite{Calabrese:2004eu}, and the holomorphic components of the energy-momemntum tensor are related by \cite{Belavin:1984vu}
\be
 T(w)=(dz/dw)^2T(z)+{c\over 12} \big(z'''z'-{3\over 2}z''^2\big)/z'^2~,
\ee
where $T(w)=-2\pi T_{ww}(w)$ has support on one of the $\alpha$ sheets and the second term on the right hand side is the Schwartzian derivative. In particular, the vacuum expectation value reads \cite{Calabrese:2004eu}
\be
  \langle T(w) \rangle_{\alpha}={c(1-(1/\alpha)^2)\over 24} \, {\ell^{2}\over (w-u)^2(w-v)^2}~,
\ee
where $\la T(z) \ra_{\mathbb{C}}=0$ has been used. Given the above general formula one can readily evaluate
\be
 {\cal P} \, \langle T(w) \rangle_\alpha= - {c\over 12} \, {\ell^{2}\over (w-u)^2(w-v)^2}~.
 \label{twist}
\ee

On the other hand, 
the modular Hamiltonian associated with a finite interval in two dimensional conformal field theory is given by \cite{Casini:2010kt}, 
\be
  K_0={2\pi\over \ell} \int_0^{\ell} dx_1\,x_1\, (x_1-\ell) T_{22}(x_1)~,
\ee
where for simplicity we chose $u=\ell$ and $v=0$. Using now
\be
\langle T(w)T(0) \rangle_{\mathbb{C}}={c\over 2 w^4}~, \quad \langle \bar T(w) T(0)  \rangle=0~,
\ee
and substituting
\be
T_{22}={1\over 2\pi} (T+\bar T)+2T_{w\bar w}~,
\ee
yields\footnote{Recall that tracelessness of the energy-momentum tensor is equivalent to $T_{w\bar w}=0$.}
\be
 \langle T(w) K_0\rangle_{\mathbb{C}}=-{c\over 12\pi} \, {\ell^{2}\over (w-u)^2(w-v)^2}~.
\ee
As expected by the correspondence (\ref{2}), this expression is in accord with (\ref{twist}).

\section{ Conclusion: further directions}
\setcounter{equation}0
\bigskip

In this paper we have suggested a certain correspondence between the correlation functions in  space with a conical defect and those defined on regular
flat spacetime. This correspondence is rather general. It is supposed to work for any singular co-dimension two surface.
However, since it involves the insertion of a modular Hamiltonian the exact form of which is not in general known the efficiency of our proposal 
reduces to the cases when the Hamiltonian can be constructed either explicitly or perturbatively. This still includes a wide class of physics situations  
and models. We conclude with listing some potentially interesting directions for further research.

\medskip

\noindent {\it 1. Higher point functions.}
In this paper we mostly considered the expectation values, or the 1-point functions, of some operators on a conical defect the calculation of which  involves
 the analysis of 2-point functions in Minkowski spacetime. This restriction is not principle and is dictated only by our desire to illustrate our correspondence
 on some relatively simple examples in which we could compare our findings with the results already existing in the literature. 
 In fact, extension of our method to higher point functions is rather straightforward. It would be interesting to compute, for instance, the two-point function
 of scalar operators on a conical defect in a conformal field theory using the exact expressions for the correlation functions in Minkowski spacetime
 found in \cite{Osborn:1993cr}. It is one of the problems which we plan to study in the future. 
 
One of the interesting related problems is to verify the precise way, conjectured in \cite{SS},  the confomal $a$-anomaly  appears in the  2-point function of energy momentum tensor considered on a conical defect. This is considered in an accompanying  paper \cite{SS1}.
 
 \medskip
 
\noindent {\it 2. Twist operators.} In \cite{Calabrese:2004eu} it was shown that computation of the field theory partition function on $\alpha$-folded cover of $\mathbb{R}^2$ is the same as the correlation function arising from the insertion of twist operators, $\Phi_\alpha$, into each of the $\alpha$ decoupled sheets. In particular, it was shown that in the two dimensional space-time, $\Phi_\alpha$ is a local primary operator with certain scaling dimension. It allows to reduce a given computation on the replicated geometry to a correlator on a regular manifold with insertion of twist operators. However, the higher dimensional counterpart of the 2D twist operator is not local. In $d>2$ it is supported on a co-dimension two entangling surface which is not point like anymore. As of today, understanding of the higher dimensional twist operators is very much limited. Our proposal (\ref{2}) suggests a possible tool for studies of $\Phi_\alpha$ in general dimension. In particular, it suggests the following operatorial identity
\be
 {\cal P}\,\Phi_\alpha=K_0~.
 \label{twist1}
\ee
This identity emphasizes that the problem of finding a twist operator and modular Hamiltonian are equivalent to certain extent. This allows us to conjecture a particular relation
\be
\Phi_\alpha=e^{(1-\alpha)K_0}\, 
 \label{twist2}
\ee
which expresses the twist operator in terms of the modular Hamiltonian. 

Yet, a word of caution about equations (\ref{twist1}) and (\ref{twist2}) should be said. While by definition the twist operators on the left hand side of these expressions are supported on a co-dimension two entangling surface, operators on the right hand side are associated with a co-dimension one submanifold. Hence, these conjectures should be taken with a big grain of salt, and we find it interesting to check our proposal in the cases when both the twist operator and the modular Hamiltonian are explicitly known. More arguments in favor of this conjecture have been given in \cite{HMS}.

\medskip

\noindent {\it 3. More general surfaces and QFTs.} The computations presented in this work are mainly focused on a planar entangling surface in Minkowski space. Hence, we find it instructive to generalize our findings by including into consideration curved geometries. Of course, the absence of known modular Hamiltonian is one of the main obstacles that must be confronted in an attempt to pursue such endeavor. Perhaps, the best point of departure would be to start these studies from spherical regions in Minkowski space, for which the modular Hamiltonian is known \cite{Casini:2010kt}, and then proceed perturbatively  to study more complicated geometries \cite{Rosenhaus:2014woa}. It would be also interesting to test our correspondence in the case of non-conformal interacting field theories such as $\phi^4$ \cite{largeN}.

\bigskip

We plan to explore these and other possible directions in the future.

\section*{Acknowledgements} 

MS is supported in part by NSF Grant PHY-1214644, and by Berkeley Center for Theoretical Physics.
SS would like to thank the Yukawa Institute for Theoretical Physics (Kyoto), especially Tadashi Takayanagi,  for hospitality
during the final stages of this project. We thank C. Bachas, P. Caputa, R. Emparan, V. Frolov, B. Pioline, V. Rosenhaus, K. Skenderis and T. Takayanagi
for valuable remarks.



\begin{thebibliography}{999}

{\frenchspacing \parskip=2mm


\bibitem{Rosenhaus:2014woa} 
  V.~Rosenhaus and M.~Smolkin,
  ``Entanglement Entropy: A Perturbative Calculation,''
  arXiv:1403.3733 [hep-th].
  
\bibitem{Deser:1983tn} 
  S.~Deser, R.~Jackiw and G.~'t Hooft,
  ``Three-Dimensional Einstein Gravity: Dynamics of Flat Space,''
  Annals Phys.\  {\bf 152}, 220 (1984),\\
  S.~Deser and R.~Jackiw,
  ``Three-Dimensional Cosmological Gravity: Dynamics of Constant Curvature,''
  Annals Phys.\  {\bf 153}, 405 (1984).


\bibitem{Som}
A.~Sommerfeld,
Proc.\ Lond.\ Math.\ Soc., {\bf 28}
, 417, (1897),\\
J.~S.~Dowker,
  ``Quantum Field Theory on a Cone,''
  J.\ Phys.\ A {\bf 10}, 115 (1977),\\
  D.~V.~Fursaev,
  ``Spectral geometry and one loop divergences on manifolds with conical singularities,''
  Phys.\ Lett.\ B {\bf 334}, 53 (1994)
  [hep-th/9405143].


\bibitem{Fursaev:1994ea} 
  D.~V.~Fursaev and S.~N.~Solodukhin,
  ``On one loop renormalization of black hole entropy,''
  Phys.\ Lett.\ B {\bf 365}, 51 (1996)
  [hep-th/9412020],\\
  D.~V.~Fursaev and S.~N.~Solodukhin,
  ``On the description of the Riemannian geometry in the presence of conical defects,''
  Phys.\ Rev.\ D {\bf 52}, 2133 (1995)
  [hep-th/9501127].

 
\bibitem{Lewkowycz:2013nqa} 
  A.~Lewkowycz and J.~Maldacena,
  ``Generalized gravitational entropy,''
  JHEP {\bf 1308}, 090 (2013)
  [arXiv:1304.4926 [hep-th]],\\
  D.~V.~Fursaev, A.~Patrushev and S.~N.~Solodukhin,
  ``Distributional Geometry of Squashed Cones,''
  Phys.\ Rev.\ D {\bf 88}, no. 4, 044054 (2013)
  [arXiv:1306.4000 [hep-th]].
 
 
 
\bibitem{kabastra} 
  J.~Bisognano and E.~H.~Wichmann,
  ``On the Duality Condition for a Hermitian Scalar Field,''
  J.\ Math.\ Phys.\  {\bf 16}, 985 (1975),\\
  J.~Bisognano and E.~H.~Wichmann,
  ``On the Duality Condition for Quantum Fields,''
  J.\ Math.\ Phys.\  {\bf 17}, 303 (1976),\\
   D.~N.~Kabat and M.~J.~Strassler,
  ``A Comment on entropy and area,''
  Phys.\ Lett.\ B {\bf 329}, 46 (1994)
  [hep-th/9401125],\\
  L.~Susskind and J.~Lindesay,
  ``An introduction to black holes, information and the string theory revolution: The holographic universe,''
  Hackensack, USA: World Scientific (2005) 183 p.,\\
  J.~H.~Cooperman and M.~A.~Luty,
  ``Renormalization of Entanglement Entropy and the Gravitational Effective Action,''
  arXiv:1302.1878 [hep-th].
  
  
 
\bibitem{SS} 
  S.~N.~Solodukhin,
  ``The a-theorem and entanglement entropy,''
  arXiv:1304.4411 [hep-th].
  
\bibitem{EE}
S.~Ryu and T.~Takayanagi,
  ``Aspects of Holographic Entanglement Entropy,''
  JHEP {\bf 0608}, 045 (2006)
  [hep-th/0605073],\\
H.~Casini and M.~Huerta,
  ``Entanglement entropy in free quantum field theory,''
  J.\ Phys.\ A {\bf 42}, 504007 (2009)
  [arXiv:0905.2562 [hep-th]],\\
  S.~N.~Solodukhin,
  ``Entanglement entropy of black holes,''
  Living Rev.\ Rel.\  {\bf 14}, 8 (2011)
  [arXiv:1104.3712 [hep-th]].
  
  
\bibitem{Bhattacharya:2012mi} 
  J.~Bhattacharya, M.~Nozaki, T.~Takayanagi and T.~Ugajin,
  ``Thermodynamical Property of Entanglement Entropy for Excited States,''
  Phys.\ Rev.\ Lett.\  {\bf 110}, no. 9, 091602 (2013)
  [arXiv:1212.1164],\\
  D.~D.~Blanco, H.~Casini, L.~-Y.~Hung and R.~C.~Myers,
  ``Relative Entropy and Holography,''
  JHEP {\bf 1308}, 060 (2013)
  [arXiv:1305.3182 [hep-th]],\\
  G.~Wong, I.~Klich, L.~A.~Pando Zayas and D.~Vaman,
  ``Entanglement Temperature and Entanglement Entropy of Excited States,''
  JHEP {\bf 1312}, 020 (2013)
  [arXiv:1305.3291 [hep-th]].
 
 
   
  
\bibitem{Iellici:1997ud} 
  D.~Iellici,
  ``Massive scalar field near a cosmic string,''
  Class.\ Quant.\ Grav.\  {\bf 14}, 3287 (1997)
  [gr-qc/9704077],\\
  V. P. Frolov and I. D. Novikov, УBlack Hole Physics: Basic Concepts And New
Developments,Ф Dordrecht, Netherlands: Kluwer Academic (1998).



\bibitem{Osborn:1993cr} 
  H.~Osborn and A.~C.~Petkou,
  ``Implications of conformal invariance in field theories for general dimensions,''
  Annals Phys.\  {\bf 231}, 311 (1994)
  [hep-th/9307010],\\
  J.~Erdmenger and H.~Osborn,
  ``Conserved currents and the energy momentum tensor in conformally invariant theories for general dimensions,''
  Nucl.\ Phys.\ B {\bf 483}, 431 (1997)
  [hep-th/9605009].
  
  
  
\bibitem{cosmic} 
  V.~P.~Frolov and E.~M.~Serebryanyi,
  ``Vacuum Polarization in the Gravitational Field of a Cosmic String,''
  Phys.\ Rev.\ D {\bf 35}, 3779 (1987),\\
  M.~R.~Brown, A.~C.~Ottewill and D.~N.~Page,
  ``Conformally Invariant Quantum Field Theory in Static Einstein Space-times,''
  Phys.\ Rev.\ D {\bf 33}, 2840 (1986),\\
  J.~S.~Dowker,
  ``Casimir Effect Around a Cone,''
  Phys.\ Rev.\ D {\bf 36}, 3095 (1987),\\
  T.~M.~Helliwell and D.~A.~Konkowski,
  ``Vacuum Fluctuations Outside Cosmic Strings,''
  Phys.\ Rev.\ D {\bf 34}, 1918 (1986),\\
  B.~Linet,
  ``Quantum Field Theory in the Space-time of a Cosmic String,''
  Phys.\ Rev.\ D {\bf 35}, 536 (1987),\\
  P.~C.~W.~Davies and V.~Sahni,
  ``Quantum Gravitational Effects Near Cosmic Strings,''
  Class.\ Quant.\ Grav.\  {\bf 5}, 1 (1988).
  
  
   

\bibitem{Cappelli:1990yc} 
  A.~Cappelli, D.~Friedan and J.~I.~Latorre,
  ``C theorem and spectral representation,''
  Nucl.\ Phys.\ B {\bf 352}, 616 (1991).

\bibitem{Calabrese:2004eu} 
  P.~Calabrese and J.~L.~Cardy,
  ``Entanglement entropy and quantum field theory,''
  J.\ Stat.\ Mech.\  {\bf 0406}, P06002 (2004)
  [hep-th/0405152].
  
  
\bibitem{Belavin:1984vu} 
  A.~A.~Belavin, A.~M.~Polyakov and A.~B.~Zamolodchikov,
  ``Infinite Conformal Symmetry in Two-Dimensional Quantum Field Theory,''
  Nucl.\ Phys.\ B {\bf 241}, 333 (1984).
  
  
\bibitem{Casini:2010kt} 
  H.~Casini and M.~Huerta,
  ``Entanglement entropy for the n-sphere,''
  Phys.\ Lett.\ B {\bf 694}, 167 (2010)
  [arXiv:1007.1813 [hep-th]],\\
  H.~Casini, M.~Huerta and R.~C.~Myers,
  ``Towards a derivation of holographic entanglement entropy,''
  JHEP {\bf 1105}, 036 (2011)
  [arXiv:1102.0440 [hep-th]].
 

\bibitem{SS1} S.~ N.~ Solodukhin, 
  ``Conformal a-charge, correlation functions and conical defects,''
  arXiv:1406.5368 [hep-th].

 \bibitem{largeN} 
 M.~A.~Metlitski, C.~ A.~Fuertes,  and S.~Sachdev,  
 "Entanglement entropy in the O(N) model,"
 Phys.\ Rev.\ B {\bf 80}, 115122 (2009)
 [arXiv:0904.4477 [cond-mat.stat-mech]],\\
  M.~P.~Hertzberg,
  ``Entanglement Entropy in Scalar Field Theory,''
  J.\ Phys.\ A {\bf 46}, 015402 (2013)
  [arXiv:1209.4646 [hep-th]].
  

\bibitem{HMS}
  L.~Y.~Hung, R.~C.~Myers and M.~Smolkin,
  ``Twist operators in higher dimensions,''
  arXiv:1407.6429 [hep-th].
 
}


\end{thebibliography}
\end{document}